\documentclass[draft]{agujournal2019}
\usepackage{url} 
\usepackage{lineno}
\usepackage[inline]{trackchanges} 
\usepackage{soul}
\usepackage{amsmath}
\usepackage{natbib}
\usepackage{url}
\usepackage{here}
\usepackage{xcolor}
\usepackage{algorithm}
\usepackage{algpseudocode}
\usepackage{soul}
\newcommand{\urlref}[2] {\href{#1}{#2}\footnote{\url{#1}, retrieved \today.}}
\usepackage{hyperref}
\hypersetup{colorlinks=true,citecolor=blue}

\draftfalse

\journalname{Journal of Advances in Modeling Earth Systems (JAMES)}

\newcommand{\order}{\ensuremath{\mathcal{O}}}

\DeclareMathOperator*{\argmin}{argmin}

\DeclareMathOperator*{\E}{E}
\DeclareMathOperator*{\Var}{Var}
\DeclareMathOperator*{\GP}{GP}

\begin{document}

%
%

\title{Semi-automatic tuning of coupled climate models with multiple intrinsic timescales: lessons learned from the Lorenz96 model}

%
%




\authors{Redouane Lguensat\affil{1}, Julie Deshayes\affil{2}, Homer Durand\affil{2}, V. Balaji\affil{3,4,5}}

\affiliation{1}{Institut Pierre-Simon Laplace, IRD, Sorbonne Universit\'e, Paris, France}
\affiliation{2}{LOCEAN-IPSL, CNRS, Sorbonne Universit\'e, Paris, France}
\affiliation{3}{Laboratoire des Sciences du Climat et de l'Environnement, CEA Saclay, Gif Sur Yvette, France}
\affiliation{4}{Princeton University, Program in Atmospheric and Oceanic Sciences, Princeton, USA}
\affiliation{5}{NOAA/Geophysical Fluid Dynamics Laboratory, Ocean and Cryosphere Division, Princeton, USA}





\correspondingauthor{Redouane Lguensat}{redouane.lguensat@ipsl.fr}




\begin{keypoints}
\item The History Matching method is explained in detail then used for tuning a toy coupled model: the Lorenz 96 model
\item The importance of several design choices is demonstrated, especially when considering forced experiments such as AMIP and OMIP
\item We argue that this tuning method is semi-automatic and highlight the importance of the expert-in-the-loop paradigm when considering it for real coupled models
\end{keypoints}

%
%

%
%


\begin{abstract}
The objective of this study is to evaluate the potential for History Matching (HM) to tune a climate system with multi-scale dynamics. By considering a toy climate model, namely, the two-scale Lorenz96 model and producing experiments in perfect-model setting, we explore in detail how several built-in choices need to be carefully tested. We also demonstrate the importance of introducing physical expertise in the range of parameters, a priori to running HM. Finally we revisit a classical procedure in climate model tuning, that consists of tuning the slow and fast components separately. By doing so in the Lorenz96 model, we illustrate the non-uniqueness of plausible parameters and highlight the specificity of metrics emerging from the coupling. This paper contributes also to bridging  the communities of uncertainty quantification, machine learning and climate modeling, by making connections between the terms used by each community for the same concept and presenting promising collaboration avenues that would benefit climate modeling research. 

\end{abstract}

\section*{Plain Language Summary}
Climate models are computer simulation codes that incorporate centuries of human knowledge of the physics of planet Earth. They are used to understand the past, the present and make projections about the future of our climate. To validate a climate model, scientists tune a number of its parameters so that it yields a simulated climate resembling real-life observations as much as possible. The main challenge in this tuning task is the extreme cost of climate models which limits a lot the number of tuning experiments scientists can run. In this paper we are interested in a technique that uses artificial intelligence in order to replace the expensive climate model with a cheaper surrogate. We experiment on a simplified model to assess the strengths and weaknesses of this semi-automatic technique, and show that it can be more efficient when combined with human expertise.

%
%

\section{Introduction}
\label{sec:intro}

Climate models, or Earth system models (ESMs), have become a primary means of exploration
of our changing climate. Numerical models of the Earth system were among the earliest
applications of digital computing \citep[see][for a participant's account of early attempts at numerical modeling of the atmosphere]{ref:platzman1979}. This soon gave rise both to numerical weather prediction, and studies of the climate, what John von Neumann called the ``infinite forecast'' \citep{ref:smagorinsky1983}, the statistics of weather fluctuations over long time periods. The inclusion of the ocean circulation into the climate system, starting with \cite{ref:manabebryan1969}, also led to our first attempts to understand the radiative and thermal balance of the planet under changes in CO$_2$ concentration \citep{ref:manabewetherald1975}. This eventually leads to the series of reports issued by the Intergovernmental Panel on Climate Change, the latest one being the sixth \footnote{\url{https://www.ipcc.ch/assessment-report/ar6/}}, where the human footprint on climate and biodiversity is now starkly visible, with fossil fuels being a primary cause.

The anatomy of an ESM consists of several model components, such as the atmosphere, ocean, and land surface. The fundamental basis of an atmosphere or ocean model consists of solving the fluid dynamics of the medium, air or seawater, with the additional complexity of the myriad processes that govern the Earth system, including water in all its phases, as well as marine and terrestrial ecosystems. Starting with the basic atmosphere-ocean general circulation model (AOGCM) of \cite{ref:manabebryan1969}, ESMs have steadily grown in complexity, adding more and more detail in terms of process representations as well as resolution (see, for instance, Figure~3.1 in \cite{ref:nasem2012}, synthesizing model evolution over several IPCC assessment reports). The processes span several orders of magnitude in time and space scales, ranging from collisions between cloud particles whose sizes are measured in microns, to deep abyssal circulations in the ocean that are the size of continental basins of \order($10^3$--$10^4$~km), where heat and carbon can accumulate over millennia. This scale range, in conjunction with complex physical, chemical and biological processes that govern the planet, make the construction of ESMs a challenging problem indeed. In the Anthropocene, this system is also subject to human influence, and thereby human choices on the ocean, atmosphere, and land surface.

There are by now about \urlref{http://esgf-ui.cmcc.it/esgf-dashboard-ui/data-archiveCMIP6.html}{116 models from 44 institutions} around the planet building ESMs to understand the climate and work out the implications of various climate policy choices. The construction of these ESMs is generally described in ``model documentation'' papers, that detail the various scientific and technical choices that went into the building of these models. We take here as examples the documentation of the most recent model versions from two of the world's premier modeling institutions, the Geophysical Fluid Dynamics Laboratory (GFDL) \citep{ref:heldetal2019}, and the Institut Pierre-Simon Laplace (IPSL) \citep{ref:boucheretal2020}. While these articles describe complete coupled models, they typically begin by describing the choices made in the ``fast'' component, the atmosphere \citep[e.g., ][]{ref:zhaoetal2018a,ref:zhaoetal2018b} for GFDL's model, \citep{ref:hourdinetal2020} for IPSL. Typically this component is first constructed, by building or refining representations of the various physical processes contributing to the atmospheric circulation, applying the most recent physical knowledge and observational constraint. A frequent method to validate the atmospheric component is to run experiments conforming to the Atmospheric Model Intercomparison Protocol \citep[AMIP:][]{ref:gates1992}. AMIP experiments typically run about 30 simulated years (SY). Subsequently this model is coupled to the ``slow'' component, the ocean. (Other components of the climate, for example changes in the biosphere are also "slow" relative to the atmosphere: in this paper, we use the ocean to stand in as the canonical "slow" component.) While the ocean component can be validated to a certain extent following the Ocean Model Intercomparison Project \citep[OMIP:][]{ref:griffiesetal2016}, several key phenomena of interest such as the El-Ni\~no Southern Oscillation (ENSO), or simply sea-ice cover in northern and southern hemispheres, are emergent properties of the coupled ocean-atmosphere system, as noted by \cite{ref:heldetal2019} and \cite{ref:adcroftetal2019} for GFDL, and \cite{ref:mignotetal2021} for IPSL.

We draw attention to the appearance of the word ``tuning'' in the title of several of the articles cited above \citep{ref:zhaoetal2018b,ref:hourdinetal2020,ref:mignotetal2021}. The tuning, or calibration as termed by statisticians, of ESMs is itself an object of study in its own right \citep{ref:hourdinetal2017,ref:schmidtetal2017}. Broadly speaking, this takes place in two stages, a first stage where we vary a few parameters in each process representation to bring them individually within observational constraints, and a second stage where the whole integrated system is calibrated to bring it within global constraints such as the Earth's radiative balance as observed from space by satellites. In the ``traditional'' approach to model calibration, described above, the model is run forward for a sufficient time for a given choice of parameters to validate it against observations. For the full coupled system including the ``slow'' physics of the ocean, doing this evaluation over a range of parameter choices can be expensive \citep{ref:adcroftetal2019}, 50,000-150,000~SY, limiting the exploration.

The stalling of arithmetic speed in computing in recent years has led to a similar slowdown in the addition of detail in ESMs \citep{ref:balaji2021}, compounding the problem. On the other hand, new computing hardware is well-suited to methods of machine learning (ML), leading to a reawakening of interest in ML methods first pioneered in the 1950s and 1960s \citep[for a recent review on the subject]{ref:sonnewaldetal2021}. 
We are interested in particular by the use of ML techniques for \textit{surrogate modeling} where the broad approach is to use ML to \emph{emulate} the expensive forward models. The emulator is then used to perform the explorations of parameter space for calibration, minimizing the number of instances of the forward model that need to be run for calibration. One recent approach applied for climate modeling is the \emph{Calibrate-Emulate-Sample} method of the CliMA group \citep{ref:clearyetal2021} where a limited set of runs of an atmospheric model is used for a broad characterization of an attractor in the parameter space of the model, by comparison with a reference, typically a model higher up ``Charney's ladder'' \citep{ref:balaji2021}, such as a large-eddy simulation (LES) model \citep{ref:dunbaretal2020,ref:couvreuxetal2020}. Such models are very high resolution and very expensive and cannot be run on climate timescales \citep{ref:schneideretal2017}. Emulators are then used to refine the landscape of the attractor, which can then be sampled to yield optimal values of parameters, as well quantify the uncertainty bounds on these values.

In an alternate approach pioneered by the HighTune Group \citep{ref:couvreuxetal2020,ref:hourdinetal2020a}, a version of the \emph{History Matching} (hereinafter noted HM) method developed by \cite{ref:williamsonetal2013} is employed. In this method, the emulator is used in successive waves not to find optimal parameter values, but to eliminate implausible regions of parameter space, according to a chosen set of metrics (distance between model outputs and observations). The forward model is then used to sample only the region of parameter space not ruled out yet (NROY). If the NROY space is a null space (i.e no plausible parameter values that satisfy constraints to within tolerance), the method identifies ``structural error'' in the model, which is unable to satisfy observational constraints \citep{ref:williamsonetal2015}.

While these methods show tremendous promise, their application so far has been in the tuning of a small component of the model, the physics of shallow clouds calibrated against an LES model where shallow cloud dynamics and physics are resolved. It has been shown by \cite{ref:hourdinetal2020a} that these methods can accelerate atmosphere model tuning relative to the conventional methods. An open question is one of how these methods will fare in the presence of the multiple timescales of the ocean-atmosphere system. The challenges posed to some pioneering ML methods for the ocean's timescales have been outlined in \cite{ref:sonnewaldetal2021}.

Some of these questions are already under investigation in a coupled model, and the challenges are becoming clear \citep{ref:hourdinetal2022}. In this article, we use a canonical idealised model of the climate, the two-scale Lorenz96 (L96, described below in Section~\ref{sec:L96}) system to explore the question of tuning in the presence of multiple timescales. The L96 system has often been used for studying climate models with a dynamical-systems lens, e.g. \cite{ref:schneideretal2017a, ref:christensenberner2019}.

In the original description by Lorenz, this model represents westward propagating waves, the slow variable, amidst a noisy background, the fast variable, as a simplification of atmospheric flows. We envision wider implications of our results for model tuning. Tuning a coupled ocean-atmosphere model requires to encompass conjointly a slow variable, the ocean, and a faster one, the atmosphere. Alternatively, tuning an ocean model alone (as forced by prescribed atmospheric conditions), requires to tune parameterizations of fast meso and submesoscale processes,  and larger-scale slower dynamics such as horizontal gyres or the meridional overturning circulation, with the risk of introducing compensating parameter errors that offset each other to provide approximately optimal solutions. These two illustrations are not unique, as one can easily find other examples of multi-scale dynamics in the climate system that require specific \emph{ad hoc} tuning to respect direct and inverse energy cascades across scales. Here we use the L96 system where we interpret the slow and fast layers of the system as ocean and atmosphere, respectively. This paper explores one of the methods for automatic tuning outlined above, the History Matching (HM) method, in the presence of multiple timescales.

The paper is organized as follows: the two-scale L96 model and  the History Matching algorithm are presented in section 2. In section 3, dedicated to the results of various experiments in a perfect model framework, we start with direct applications of HM to the full L96 model, and then consider  AMIP and OMIP-like experiments i.e.  experiments where only one of the two variables is explicitly resolved. We finally discuss,  in Section 4,  the lessons learned from applying HM to L96 and the open research avenues that could lead to an efficient application of HM unto coupled climate models.

\section{Materials and Methods}
\label{sec:methods}

\subsection{The L96 model}
\label{sec:L96}

Introduced by Edward Lorenz in a ECMWF workshop on predictability \citep{lorenz1996predictability}, the L96 model is still one of the most used toy models in geosciences. It serves as a simple test bed to investigate the performance of algorithms related to dynamical system forecasting, parameterization, data assimilation and more \citep{lorenz1998optimal, lguensat2017analog, rasp2019online, gagne2020machine, chattopadhyay2020data}.

The two-scale Lorenz-96 model consists of $K$ slow variables $X_k(k = 1, ... , K)$, each of which is coupled to $J$ fast variables $Y_{j,k}(j = 1, ... , J)$. The ODE system is described by the following equations:

\begin{gather} \label{eq:l96}
   \frac{d X_{k}}{dt}=\underbrace{-X_{k-1}\left(X_{k-2}-X_{k+1}\right)}_{\text {Advection }} \underbrace{-X_{k}}_{\text {Diffusion}}\underbrace{+ F}_{Forcing} \underbrace{- \frac{hc}{b} \sum_{j=1}^{J} Y_{j, k}}_{Coupling} \\
    \frac{d Y_{j, k}}{d t}=\underbrace{-cb Y_{j+1, k}\left(Y_{j+2, k}-Y_{j-1, k}\right)}_{\text {Advection }} \underbrace{-cY_{j, k}}_{\text {Diffusion }} \underbrace{+\frac{hc}{b} X_{k}}_{\text {Coupling }}\label{eq:l96Y}
\end{gather}

\begin{figure}[!t]
  \begin{center}
    \includegraphics[width=0.5\textwidth]{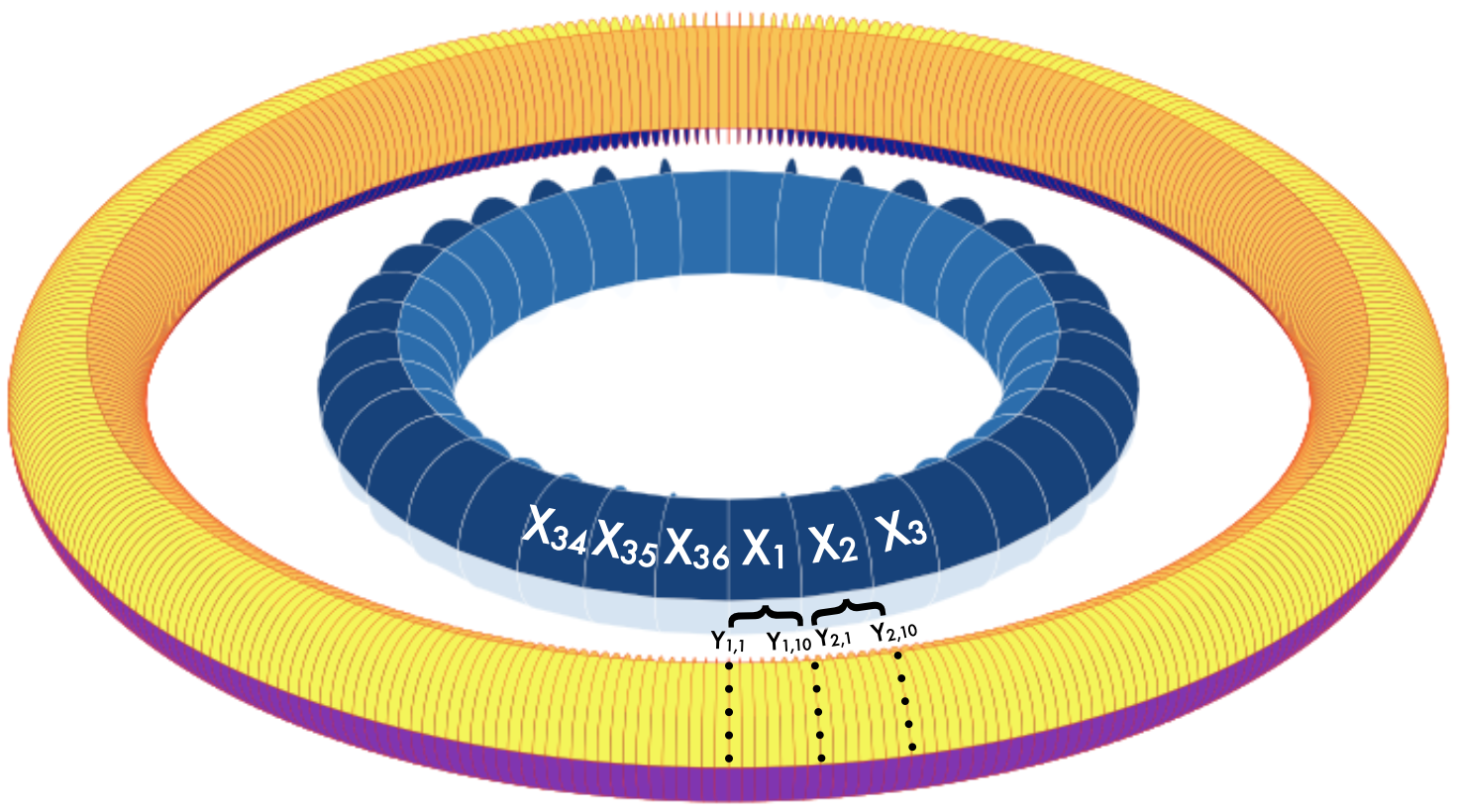}
    \includegraphics[width=0.4\textwidth]{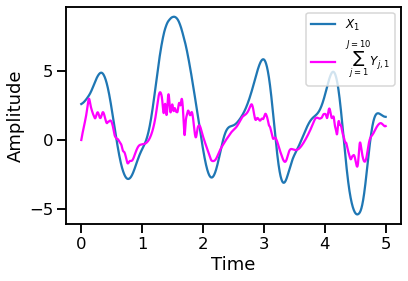}
  \end{center}
  \caption{The two-scale Lorenz 96 system consists of a non-linear ``slow'' ODE of $X$ variables coupled to a non-linear ``fast'' ODE of the $Y$ variables. In our case, the number of $X$ is 36, where each X is coupled to 10 $Y$ variables as shown in this schematic. Time is expressed in MTUs (Model Time Units) where 1 MTU $= 1000 \Delta t$}
  \label{fig:l96}
\end{figure}

The two types of variables have periodic boundary conditions and are arranged in a cyclic way ( Fig.~\ref{fig:l96}). The system is integrated using a Runge–Kutta fourth order scheme with a time step of $\Delta t = 0.001$ (note that L96 is fully non-dimensional). We based our implementation of the L96 on the Python code accompanying the paper of \cite{rasp2019online} \url{https://github.com/raspstephan/Lorenz-Online}, with the difference being in using the original formulation of the L96 as stated in Eq. (1,2).

As in Lorenz' seminal paper, we use K = 36, J = 10, a chaotic behavior is ensured by using the parameters h = 1 and F = c = b = 10 \citep{lorenz1996predictability}. In such setting, the temporal-scale ratio $c$ implies that the Y (fast) variables fluctuate ten times rapidly as the X (slow) variables, and $b$ the spatio-scale ratio implies that their amplitude is 1/10 of the slow variables.

As noted above in Section~\ref{sec:intro}, this article intends to reproduce the process of tuning of a system with multiple temporal scales, in order to help us anticipate what might transpire when the HM method is applied to a full coupled climate model. We underline that this is a gross simplification, as the real system has many interlocking feedbacks and more than just two timescales. The forcing $F$ is also stationary here, whereas the external forcing in the real system has both short-lived (e.g. volcanic eruptions) and slowly-evolving (e.g. greenhouse gas emissions) components. Nevertheless, we will demonstrate that there are important lessons to be learned regarding the coupling of multiscale systems, and this article will walk through the steps of automatic tuning to illustrate the strengths of the objective approach, as well as aspects requiring caution.

Concretely, we use time-averaged statistics of the variables produced by such a system, with the settings mentioned above, as observations and consider the History Matching technique as a means to tune the four parameters $\{F,h,c,b\}$. As this is an abstract model with no observational counterpart  we set ourselves in a \emph{perfect model} framework, and consider a preliminary realization of the L96 model as the trajectory from which we can extract analog of the climatological quantities used to tune climate models, i.e. the ``truth''. Note that we are interested in approximating the model's \emph{climate} (attractors in its parameter landscape), not the actual trajectory itself (its ``weather''). Afterwards, we pretend that we do not know which values of the parameters is the most appropriate to replicate the truth and examine to what extent the History Matching technique is capable of finding values of parameters $\{F,h,c,b\}$ that produce a suitable approximation of the true model climate.

\subsection{The History Matching method}
\label{sec:HM}

The term "History Matching" (HM) stems from the oil engineering community, where the use of statistical "reservoir simulators" instead of complex and expensive fluid flow models serves to find inputs (reservoir geology for example) for which the outputs closely match historical hydrocarbon reservoir production \citep{craig1997pressure, sacks1989design}. HM is a well published and established method used in several science and engineering applications such as galaxy formation models \citep{bower2010galaxy} and infectious disease models \citep{andrianakis2015bayesian}, to name a few.

\begin{figure}[!t]
    \includegraphics[scale=0.5]{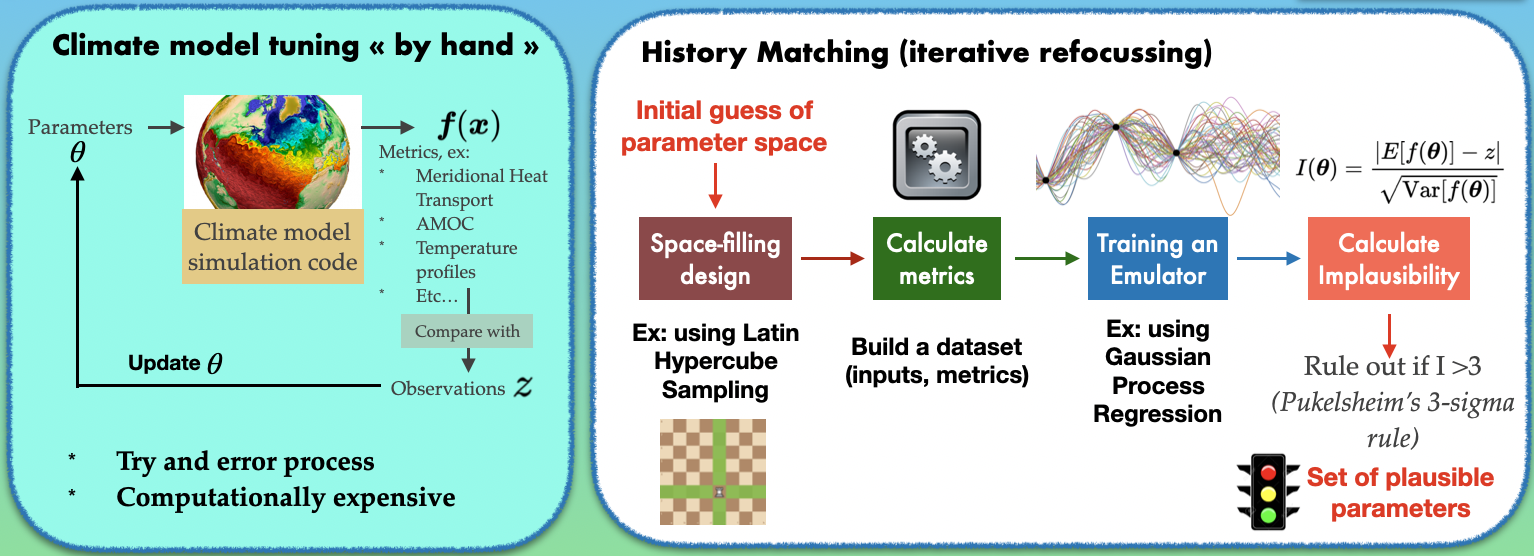}
    \caption{Manual model tuning vs. History matching}
    \label{fig:HM}
\end{figure}

The uptake of these methods in climate model development is perhaps not as high as it could have been, and it is still an open question whether this is simply due to conservatism, or whether there are indeed good reasons why these methods do not scale to full-complexity models \citep{salter2019uncertainty}. However there are recent forays into the use of HM in atmospheric model development, as noted above in Section~\ref{sec:intro}, notably via the HighTune project \citep{ref:couvreuxetal2020,ref:hourdinetal2020, villefranque2021process}. A schematic comparison of ``conventional'' methods with the HM method is shown in Figure~\ref{fig:HM} where the modeler begins with a certain number of parameters imperfectly constrained by observations. In the conventional method, this parameter space is explored by trial and error, where a number of (expensive) runs of the forward model are examined by experts to arrive an at ``optimal'' set of parameter values. In the HM approach, few runs of the expensive model are used to construct a  ``training'' dataset used by fast emulators or surrogates such as Gaussian processes \citep{kennedy2001bayesian, gramacy2020surrogates} which mimic the parameterized model and extrapolate on the whole parameter space. 
Successive ``waves'' iteratively eliminate implausible regions of parameter space, and the not-ruled-out-yet (NROY) space presents a small region of good set of parameters to be explored, using the full (expensive) model rather than a surrogate.

In the machine learning community, the process of using cheap emulators with the aim of derivative-free calibrating of costly black-box models, belongs to the field of surrogate modeling or meta-modeling  \citep{gramacy2020surrogates}. HM thus benefits from many of the advancements in this field, and is affected by the ongoing revolution in machine learning techniques and tools.

It is important to note that many similarities arise when comparing Bayesian calibration and HM, they however have fundamental differences in their strategies. In fact, calibration by construction will always result in a posterior distribution over all the input space (because the distribution needs to sum up to one) and then will always find a solution to the calibration problem. History matching, in contrast, focuses on ruling out bad solutions, and might even in some cases lead to an empty set of solutions. In Section 6, we contrast this approach with more traditional parameter optimization methods in the presence of structural and epistemic uncertainty. This article aims to highlight the advantages, from a climate perspective, of finding a bounded region of parameter space where any value adequately fits our observational bounds, as opposed to ``optimal'' values.

Algorithm \ref{alg:hm} summarizes the overall steps of the HM method, which are detailed in the following subsections.

\begin{algorithm}[!t]
\caption{History Matching with iterative refocusing}\label{alg:hm}
\begin{algorithmic}
\Require the (expensive) simulator $S$ + a priori information about the range of $P$ parameters + $M$ predefined metrics of interest 
\State \textbf{Step 1: Select a number of parameter sets}
    \begin{itemize}
    \item Define the hypercube formed by the  ranges of the $P$ parameters.
    \item Sample $N=10 \times P$ training points from that parameter space using maximin Latin Hypercube Sampling (or another space-filling design method).
    \end{itemize}
\State \textbf{Step 2: Calculate the metrics}
    \begin{itemize}
        \item Evaluate the simulator at each of the $N$ points. Construct the training data $D_{Train}=\left\{\mathbf{p}_{n}, S(\mathbf{p_n})\right\}_{n=1}^{N}$.
    \end{itemize}    
\State \textbf{Step 3: Train the emulators}
\begin{itemize}
    \item Train a GP-based (or another ML-based) emulator for each of the $M$ metrics (following a single GP per single output strategy).
\end{itemize}
\State \textbf{Step 4: Calculate Implausibilities}
\begin{itemize}
    \item Evaluate the implausibility  $I(\mathbf{p}')$ over a large number (usually hundreds of thousands) of parameter samples $\mathbf{p}'$ from the initial hypercube, using the emulators. 
    \item Identify NROY space as points where the implausibility is less than 3 for each of the $M$ metrics (conservative strategy). Alternatively, a finite number of metrics may be allowed to have implausibility higher than 3.
\end{itemize}
\State \textbf{Step 5: Iterative refocusing}
\begin{itemize}
    \item Resample the NROY space and repeat steps $2-4$ unless
    \begin{itemize}
        \item the emulator's uncertainty $V_{c}$ is smaller than the other uncertainties,
        \item NROY is empty (all space is implausible),
        \item the computational allowance for running the simulator has exhausted.
    \end{itemize}
\end{itemize}
\end{algorithmic}
\end{algorithm}

\subsubsection{Space filling design}

The need for collecting small informative samples that are cheap and at the same time permit a good representation of a whole search space, made sampling methods at the heart of Design of Experiments (DoE or DoX), a research area on its own. 

In surrogate modeling in general and HM in particular, starting from a guess interval for the set of parameters 
, we attempt to “fill” parameter space as uniformly as possible, and, if possible, aim for minimal correlation between the parameters in the design. Two families of sampling methods exist in the literature: space-filling designs and low-discrepancy sequences  \citep[see][for a review of these methods]{steponavivce2016sampling}. As in most recent literature and in particular in \cite{williamson2017tuning}, we use a popular space-filling design technique: maximin Latin Hypercubes Sampling (LHS) \citep{mckay2000comparison, morris1995exploratory}. Note that the space of ranges of all parameters describes a hyper-rectangle rather than a hypercube, until the parameters are normalized, a procedure applied later in the algorithm to help training the GP.

 \cite{chapman1994arctic}, a study from the sea ice modeling community, introduced the rule of thumb to determine the number of samples, and suggested that a number of $ 10 \times P$ samples where $P$ is the number of model parameters, is usually a good starting point. \cite{loeppky2009choosing} ran a number of experiments to further support this rule of thumb. We adopt this strategy in our experiments, meaning that to tune $\{h, F, c, b\}$ we run 40 simulations at each HM wave.

\subsubsection{The metrics}

\textit{Metrics} here refer to the quantities chosen by the modeler to assess the performance of the model. Their role in tuning is crucial since they constrain the parameter search. We note that in the machine learning community, the term "metrics" refers usually to mathematical functions used for evaluating the performance of machine learning models (such as MSE, cross-entropy, etc.). Here it is the performance of black-box simulator, $S$, i.e. the climate model in practice, that the metrics are expected to assess. 

When tuning a climate model, the metrics are physical quantities that reflect the distance between climate model outputs and observations \citep[for example the shortwave cloud radiative effect as in][]{ref:hourdinetal2020a}. As the L96 model is only a toy version of a climate model, we chose as metrics the first and second order momenta of both variables, as well as their covariances :

\begin{equation} \label{eq:metrics}
\boldsymbol{f}(X, Y)=\left(\begin{array}{c}
\langle X\rangle_{T} \\
\langle \bar{Y} \rangle_{T} \\
\langle X^{2}\rangle_{T} \\
\langle X \bar{Y}\rangle_{T}\\
\langle \bar{Y^{2}}\rangle_{T}
\end{array}\right)
\end{equation}

\noindent where $X$ represents the K=36 $X_k$ variables, and $\bar{Y}$ represents the means of the $J=10$ $Y_{k,j}$ variables associated to each $X_k$, hence the metrics vector $f(X,Y)$ includes $36*5=180$ metrics. Note that the carets denote the time average of a function $\phi(t)$ over the time interval $[t_0, t_0+ T]$,
$$
\langle\phi\rangle_{T}=\frac{1}{T} \int_{t_{0}}^{t_{0}+T} \phi(t) \mathrm{d}t,
$$
which we calculate using discrete sums instead of integrals. 

\subsubsection{The emulator}

In the surrogate modeling literature, Gaussian Processes (GP) have been the method of choice when dealing with tuning or calibration problems. GP are simple regressors that have the advantage of yielding not only mean estimates but also uncertainties.
As in \cite{williamson2017tuning}, GP are considered in this work as a viable candidate to emulate the metrics in Eq.(3).

Let us denote $p$, a set of the L96 four parameters as a \emph{configuration}, and $f(p)$ the vector of the metrics associated with the integration of $X$ and $Y$ variables using this configuration of L96. The mathematical model used for the regression is based on the following formula:

\begin{align}
    f_i(p) &= \sum_j \beta_{ij}g_j(p) + \epsilon_i(p) \\
    \epsilon_i(p) &\sim \GP(0, C_i(., .; \phi_i))
\end{align}

\noindent where $g(p)$ contains a library of basis functions in $p$ (polynomials for example), $\beta$ is a set of trainable coefficients, $C_i$ denotes pre-specified covariance functions for the GP and $\phi_i$ are their parameters. Note that the two equations can be summarized into one by taking the linear regression part as a mean function of the GP. Training the overall regression model resorts to fitting all the parameters of the GP. 

In this work, we take a blind approach to the fitting of the mean function, i.e. no human expertise is available on the choice of basis functions. We however note that if such expertise is available it must be prioritized. The regression can then be performed using classical ordinary least squares or sparsity enforcing techniques such as LASSO ; here we follow instead a forwards and backwards stepwise regression \citep{draper1998applied} approach. Regarding the covariance function, again several choices are available (Radial Basis Function (RBF), Matérn, Periodic, etc.) but throughout this paper we use the RBF kernel for its simplicity. 

For the validation of our GP-based emulators we use a Leave-One-Out approach (L1O), since we work in a small dataset setting (tens of samples). We consider the emulator to be valid if more than $80$\% of L1O prediction intervals contain the true samples.

\subsubsection{The implausibility}

A natural way to find the configuration (ie the set of parameters $p$) that allows the model to get as close as possible to the real state of the climate system, resorts to defining a distance measure between the model output $f(p)$ and the observations of the real system $z$. Note that these observations are eventually biased, due to instrumental inaccuracies for example, hence the metrics calculated from observations are eventually distinct from the actual true metrics, noted $y$ below. 

Using the emulator to find the configuration that minimizes the distance between the model output and the system state, comes to solving the following optimization problem  :
$$ p^* = \argmin_p \|z - f(p)\|$$
where $\|.\|$ is a norm taking into account different uncertainty sources. Here we use the Mahalanobis distance
$$  \|z-f(p)\|=(z-f(p))^{T} \operatorname{Var}[z-f(p)]^{-1}(z-f(p))$$

Because we only run a selected number of simulations and use the emulators to complete the hypercube of parameters, we do not have access to entire distribution of the model $f(p)$ but only to the expectation $\operatorname{E}[f(p)]$ and to the variance $\operatorname{Var}[f(p)]$. Following notations in \cite{williamson2017tuning}, we denote $V_e$ the error variance of observations and $V_\eta$ the error variance due to the simulator uncertainties. We can then reformulate the distance to observations, using the prediction of the emulator, and we refer to it as the \textit{implausibility} :
\begin{align*}
    \mathit{I}(p) &= \|z-\operatorname{E}[f(p^*)]\\
    &= (z-\operatorname{E}[f(p^*)])^{T}\| \operatorname{Var}[z-\operatorname{E}[f(p^*)]]^{-1}(z-\operatorname{E}[f(p^*)]) \\
    &= (z - m^*(p^*))^T\Var[(z-y) + (y-f(p^*)) + f(p^*)-\E[f(p^*)]]^{-1}(z-m^*(p^*))\\
    &= (z-m^*(p^*))^T(V_e + V_\eta + \Var[f(p^*)])^{-1}(z-m^*(p^*))
\end{align*}   

With this definition, we factor in the implausibility the different uncertainties associated with the model itself (L96 in our case, the climate model otherwise), with observations and with the predictions of the emulators. Thus the small values of implausibility appear in two cases only: the distance between the prediction of the model and the real state of the system is small or one of the three uncertainties is too high. 

Having calculated the implausibility for a given set of parameter $p$, deciding to keep it or rule it out is based on a threshold value $T$ for the implausibility i.e. $p$ is ruled out if $\mathit{I}(p) > T$ and the NROY region is $\{p : \mathit{I}(p) \leq T\}$. For $d$-dimensional $\mathbf{z}$, \cite{bower2010galaxy} use $T=\chi_{d, 0.995}^{2}$, the $99.5$ th percentile of the $\chi^{2}$ distribution with $d$ degrees of freedom.

If $z$ is one-dimensional, as will be considered in this work, the implausibility is written as :
\begin{equation}
   I(\mathbf{p})=\frac{\left|z-\mathrm{E}\left[f(\mathbf{p})\right]\right|}{\left[V_{e}+V_{\eta}+\Var[f(\mathbf{p})]\right]^{1 / 2}} 
\end{equation}
\noindent and the Pukelsheim rule is generally used and sets $T=3$. This general rule states that any continuous unimodal distribution contains at least 95\% of its probability mass within a distance of 3 standard deviations from its mean.





\subsubsection{NROY plots}

\begin{figure}[H]
    \centering
    \includegraphics[scale=0.37]{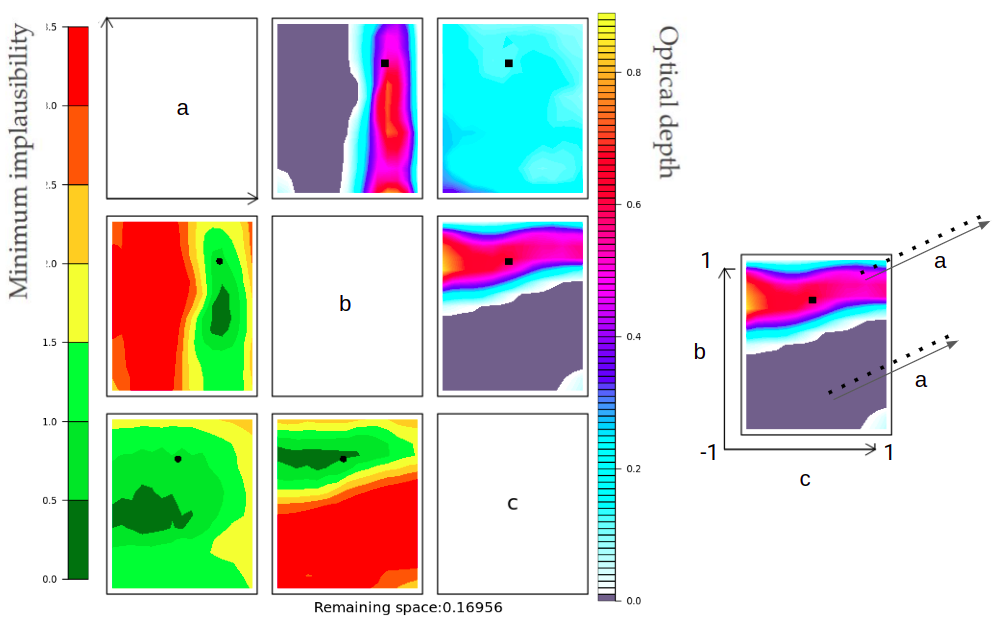}
    \caption{NROY plots for the theoretical case of applying HM to tune a 3-parameter model. Colors associated to the left-hand-side bar indicate the minimum implausibility, while those associated to the right-hand-side bar show the optical depth, for every grid point of  each 2-dimensional subplot. Subset to the right hand side provides further details on how to read each subplot.}
    \label{fig:exp_NROY}
\end{figure}

In order to visualize the NROY space, two types of information are usually considered in HM literature: \textit{minimum implausibility} and \textit{optical depth}. An example using 3 tunable parameters (a,b,c) is shown in Fig. \ref{fig:exp_NROY}, where the minimum implausibility is represented below the diagonal of the NROY ``matrix" while the optical depth is shown above it. Before explaining these two information in detail, we note that given the difficulty of visualizing a $d-$dimensional space with $d>3$, the NROY space visualization is done using a composition of $\frac{d(d-1)}{2}$ two-dimensional subplots. 

Let us consider one subplot such as the one shown to the right of Fig. \ref{fig:exp_NROY}. The ordinate and abscissa represent respectively $b$ and $c$ parameter intervals projected into $[-1, 1]$, a normalization that is done to help  training  the GP. We remind the reader that the NROY plots are obtained at test time (in ML terminology), rather than at training time. In fact, we apply our trained GP to a large variety of configurations, usually in the order of hundreds of thousands of parameter sets (in this work we initially use one million samples from a random LHS of the search space for parameters). Since the 2-dimensional subplots are restricted to combinations of 2 variables, each subplot is divided into bins in the 2 directions of the plot. In the aforementioned example, each bin has a restricted space for $b$ and $c$ but contains all the possible values of $a$. The latter group of configurations is then qualified by the minimum implausibility (i.e. the minimum implausibility amongst all configurations of that group) and the optical depth, that is the fraction of configurations with implausibility smaller than the pre-defined threshold. 
The rationale behind the minimum implausibility is that if the minimum is already higher than the threshold considered for implausibility (here 3), then whatever the choice for the other parameter(s), the metrics would be far from observations.

For easier visualization, the orientation of all subplots follow the arrows of the top-left one. The black dots in each subplot represent the ground-truth configuration, if already known.

\subsubsection{Iterative refocusing}
Once the first NROY space is produced, it seems obvious to apply the HM procedure, on this reduced space. This has the advantage of training better GP-based surrogates, because the parameter space is smaller than before, therefore reevaluating the implausibilities. We refer to this task by the term \textit{refocusing} \citep{williamson2017tuning}, and each iteration of steps 2-4 is called a wave. 

The iterative aspect of the HM provides a certain flexibility that other approaches may not. After having significantly reduced the parameter space with a set of metrics describing well the general tendencies of the system, one could  try to reduce it further by using metrics describing some more local aspects. Notwithstanding, the multi-wave HM comes with some challenges. Many flavors of HM exist in the literature, where some specific choices are made for some steps, notably, the stopping criteria and the NROY resampling strategy. Overall, the stopping criteria are problem-dependent, and the limitation of computational resources most often come to play. In a pragmatic manner, the iterative process must be halted when it is expected that performing one additional wave would not reduce the parameter space sufficiently compared to the computational cost that it would require (as a reminder, the simulator is run at each step 2 of the algorithm). Also, as stated in \citet{williamson2017tuning}, "\textit{when the  emulator variance is largely smaller than the denominator  in the implausibility calculation,  then  it is  unlikely  that  further  waves  will  change  the  implausibility  very  much}" and it may be unreasonable to perform a new wave.

In addition, a difficulty arises with multi-wave design right after the first wave :  LHS cannot sample the NROY space as it is in general not a hyper-rectangle anymore and may contain several disconnected regions. Several strategies were followed in the literature: \citet{yeh2016practical} used clustering to find medoids that act as representatives of each cluster, \citet{andrianakis2015bayesian} used Gaussian random variables centered at the mean from NROY points to sample design points for the subsequent wave. This is an open research question and recent advances in machine learning can lead to promising avenues \citep{garbuno2020history}. In the present application of HM to L96, we follow a rejection sampling approach to sample subsequent waves' NROY as in reference papers employing HM, such as \cite{bower2010galaxy} : we apply LHS on the entire initial parameter space with enough samples to retain approximately the desired $10\times P$ configurations after rejecting those with an implausibility score greater than 3 for each emulator. 


\subsubsection{Dimensionality Reduction}

A major difference in our application of HM from \cite{williamson2017tuning} and \cite{ref:hourdinetal2020a} relies on the dimension of $f(X,Y)$, the metrics vector. With as many as 180 metrics (and presumably as soon as $M>10$), it becomes relevant to reduce the dimensionality of the metrics vector using for example Principal Component Analysis (hereafter PCA, also known as Empirical Orthogonal Functions in the geoscience literature). As a consequence, the GP will be fitted during step 3 in the reduced space and not in the $M$-dimensional space. Observations are also projected in the reduced space before calculating the implausibilities. 

Reducing the dimensionality of metrics seems to be relevant for metrics that are not scalar, for example geographical maps of geophysical variables. However, due to the simplicity of L96 model, we could not test such procedure in the present study.

\subsubsection{The code}

As stated before, the L96 model is written in Python ; special attention was given to parallel computing in order to run several experiments efficiently, the \textsc{joblib} Python library was used for this task. The main HM routine is written in R (LHS, GP-fitting, L1O, visualization). The code is based on the \textsc{ExeterUQ\_MOGP} library \footnote{\url{https://github.com/BayesExeter/ExeterUQ_MOGP}} that acts as an R interface for the Python written \textsc{mogp\_emulator} library for training GPs, also available online\footnote{ \url{https://github.com/alan-turing-institute/mogp_emulator/}}. Again, to speed-up the calculation of implausibilities, we make use of the \textsc{future} R library for parallel computation. Please note that the code requires minimal R knowledge, which would not constitute a barrier for pure Python practitioners. Jupyter notebooks used for this work can be found in \url{https://github.com/HRMES-MOPGA/L96HistoryMatching}. 


\section{Results}
\label{sec:experiments}

In this section we demonstrate the relevance of the HM algorithm in a perfect model setting i.e. $(V_e = V_\eta = 0)$. Our aim is to outline new insights regarding the influence of several design choices when applying HM and to evaluate its applicability to models with different time scales. The general goal of this study is to answer several questions about the choices usually made when applying HM for climate modeling applications. Specifically, for the current paper, we aim to "history match" the metrics in Eq. \ref{eq:metrics} to reduce the search space for the four L96 parameters \{h, F, c, b\}.

\subsection{Data Generation}
\label{sec:datagen}

Using the "true" set of parameters $p_{true} = \{10,1,10,10\}$ we run a short simulation (10 MTUs) to reach a state in the L96 attractor, save that state, then run again a longer simulation starting from the attractor for 100 MTUs. This latter simulation yields a specific trajectory of the L96 model that we will consider as the ground truth for our tuning experiments. As written above, in our chosen set-up, evaluating our metrics against the latter ground truth, yields a vector of 180 dimensions (Eq.\ref{eq:metrics}) for every set of parameters $p$.

For the maximin LHS, we generate $10 \times 4 \left(\text{ie the number  parameters to tune}\right) = 40$ set of parameters for the first wave. In subsequent waves we proceed as follows: after wave number $\mathit{w}$, we estimate the search space reduction ratio $$r = \frac{\text{NROY}_w \text{volume}}{\text{Initial search space volume}}$$. We then generate a maximin LHS of size $\lceil \frac{ 10 \times 4 }{r}\rceil$, where $\lceil . \rceil$ is the ceiling function. If $r$ becomes too small (which happens naturally when HM effectively reduces a lot the search space), the maximin LHS becomes computationally heavy and we then switch to a random LHS (without maximin optimization). 

The emulation training problem calls for the use of multi-input multi-output GP (inputs: 4 dimensional, outputs: 180 dimensional). It is common to find in the literature the use of multiple single-output GP instead (one per output so 180 single-output GP in our case) due to their better scaling with memory and computational time. Obviously, this would disregard correlations between the outputs and may lead to lower performance since some information is discarded. A straightforward way to tackle both problems is through the use of dimensionality reduction algorithms. In this study, we use Principal Component Analysis and keep a number of principal components that retain 99\% of the total variance. Emulating the metrics in the reduced space with independent GP often performs as well as using a multi-output emulator, especially if the size of the training samples is large compared to the dimension of the reduced space \citep{wilkinson2010bayesian}.

\subsection{History Matching with no physical prior}
\label{sec:nophys}

We start our experiments by a standard setup where we suppose we have access to the simulator, here the L96 code, and to the ground truth output metrics, but have uninformed priors on the parameters. Hence we start with arbitrary uniform priors  except for parameter $c$ that is positive because we know from the equations that $c$ represents a time-scale ratio (Table \ref{table1}).

\begin{table}[H]
\caption{Prior intervals for the parameters along with the ground truth values, for the HM experiment with no physical based prior.}
 \centering
\begin{tabular}{c c c}
 \hline
 Params & Prior & True  \\
 \hline
F & {[-20,20]} & 10 \\
h & {[-2,2]} & 1 \\
c & {[0,20]} & 10 \\
b & {[-20,20]} & 10 \\
\hline
\end{tabular}
\label{table1}
\end{table}

\begin{figure}
    \centering
    \includegraphics[scale=0.37]{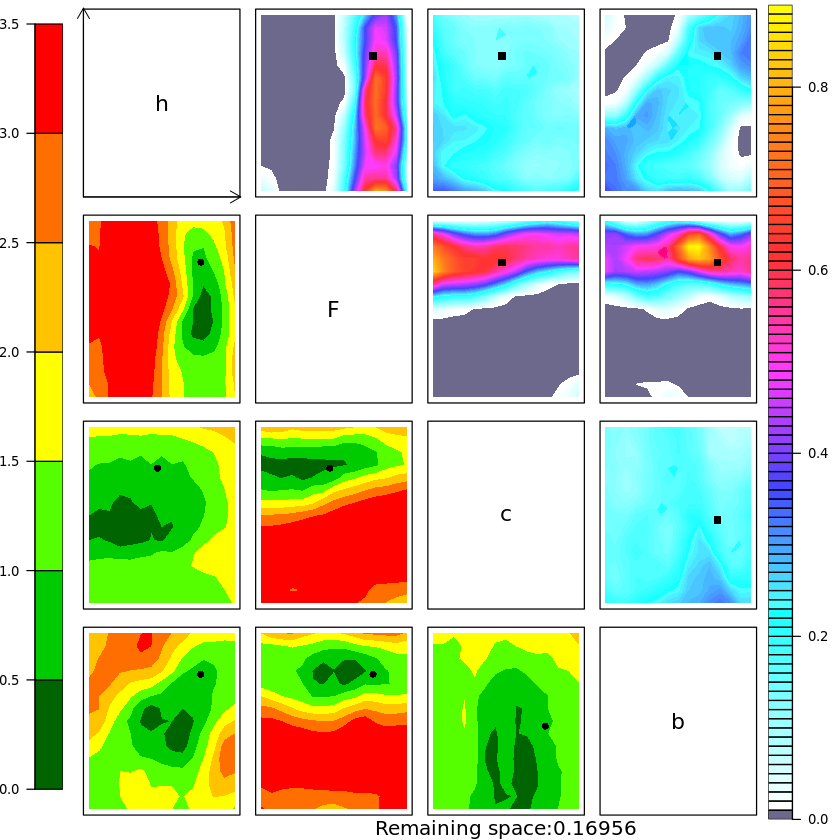}
    \includegraphics[scale=0.37]{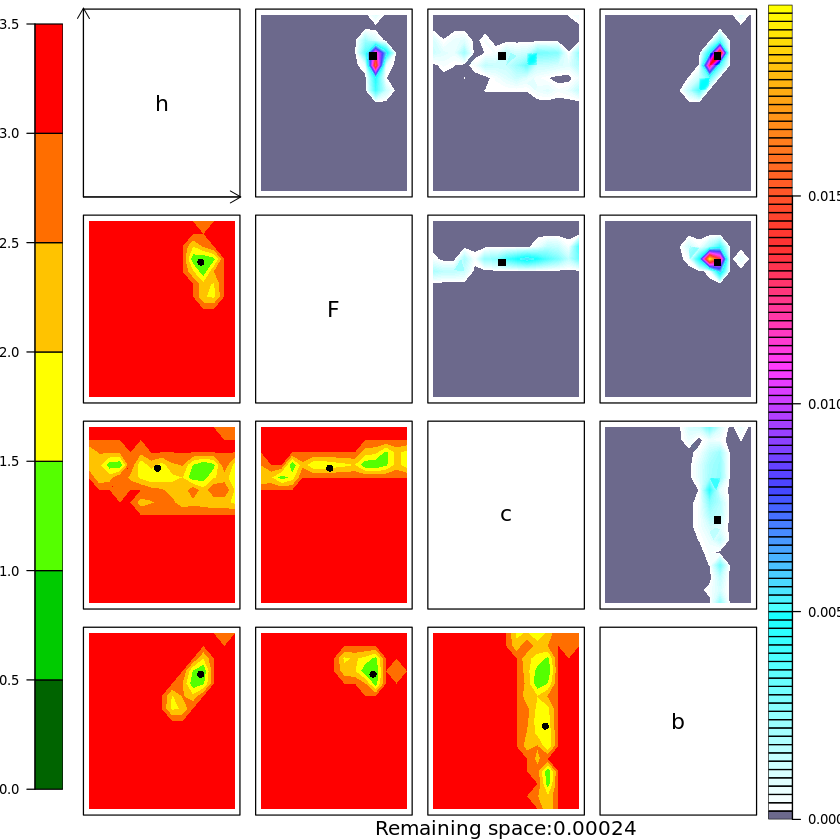}
    \caption{NROY plots for the HM experiment with no physical based prior, at wave 1 (left hand side) and at wave 6 (right hand side).}
    \label{fig:HM_noprior}
\end{figure}

We perform HM as described in the previous sections, and carry out a total of 6 waves which sums up to 240 simulations of the L96 model (Table \ref{tableNROY1}). The NROY ratio $r$ is very small, including at wave 1, which indicates that our priors are too large. More specifically, the NROY matrix suggests that most of the reduction of the parameter space is tied to the $F$ parameter (Figure \ref{fig:HM_noprior}) : nearly all the negative values are ruled-out (remember that the NROY matrix shows parameters normalized on the $[-1,1]$ interval, hence projecting back to the original space, this means that values of $F$ on the interval $[-20,0]$ are ruled-out).

\begin{table}[h]
\caption{Evolution of the NROY ratio $r$ for each wave of the HM experiment with no physical based prior. Right hand side column specifies the number of explicit simulation with L96 model carried out at each wave.}
\centering
\begin{tabular}{rrr}
  \hline
wave & NROY (\%) & NbSim \\ 
  \hline
1 & 16.96 & 40 \\ 
  2 & 7.91 & 40 \\ 
  3 & 5.55 & 40 \\ 
  4 & 2.31 & 40 \\ 
  5 & 0.94 & 40 \\ 
  6 & 0.02 & 40 \\ 
  \hline & & Total = 240 \\
  \hline
\end{tabular}
\label{tableNROY1}
\end{table}

At wave 6, we are left with $0.02$\% of the initial search space, which represents  nearly 2000 configurations (as a reminder, the initial search space contains one million samples from a random LHS). The associated NROY plots show that we successfully narrowed the search space for parameters ${h,b,F}$, but interestingly parameter $c$ presents multiple NROY regions (Figure \ref{fig:HM_noprior}, right hand side). This suggests possible metastability phenomena on time scales longer than the integration time used to calculate the metrics, which is also encountered in the Ensemble Kalman Inversion (EKI) based calibration used in \cite{ref:schneideretal2017}. It is worth highlighting again that the goal of HM is not finding \textit{exactly} the ground truth solution, but ruling out the wrong ones. Thus, we consider a HM experiment to be successful if: (1) the ground truth solution is not ruled out, (2) the parameter search space is largely reduced. For this experiment with no physical prior (other than on the sign of $c$), running additional waves does not change the NROY space further, and becomes tedious because of the rejection sampling of an extremely small region.

From the remaining configurations in the NROY of wave 6, we apply the K-means algorithm and find six clusters (according to the silhouette score (\ref{silhou}) which centers are considered as possible configurations. Before running the L96 model (our "expensive" simulator in this idealized study) with these sets of parameters, we verify that they belong to the last NROY effectively : it is not granted that a k-means cluster center is included in the NROY. Indeed, one of the centers does not respect this condition hence we finally end up with 5 configurations (Table \ref{tableNROY1final}).

\begin{table}[H]
\caption{Final sets of parameters identified by HM in experiment with no physical based prior, alongside the ground truth values (bottom line), and the corresponding median KL-div between their simulations and observations (right hand side column). } 
\centering
\begin{tabular}{rrrrr|c}
  \hline
 & h & F & c & b & KL-div\\ 
  \hline
1 & 0.99 & 11.90 & 16.21 & 9.17 & 0.13 \\ 
  2 & 1.15 & 8.94 & 3.94 & 10.33 & 0.09 \\ 
  3 & 0.58 & 10.31 & 11.90 & 6.89 & 0.15 \\ 
\textbf{4} & \textbf{0.92} & \textbf{10.31} &\textbf{ 10.28} & \textbf{9.35 }& \textbf{ 0.01} \\ 
  5 & 1.87 & 11.52 & 19.03 & 15.65 & - \\
  \hline
  ground truth & 1 & 10 & 10 & 10 & \\
   \hline
\end{tabular}
\label{tableNROY1final}
\end{table}

Finally, we run L96 simulations with the 5 sets of parameters identified by the HM procedure (Table \ref{tableNROY1final}), calculate the metrics and overlay their distribution onto that of observed metrics (i.e. our initial L96 simulation, considered as ground truth). Note that we evaluate the realism of metrics with statistical distributions, similar to climate modelers evaluating simulated climate against observations. Looking at these distributions, one configuration clearly stands out as a bad candidate (Figure \ref{fig:configs_noprior1}), although it was not ruled out by HM ! This configuration is an example illustrating that HM can retain a low-implausibility configuration not because the metrics are close to observations (low nominator of implausibility), but because the GP were uncertain (high denominator of implausibility).

\begin{figure}[H]
    \centering
    \includegraphics[scale=0.35]{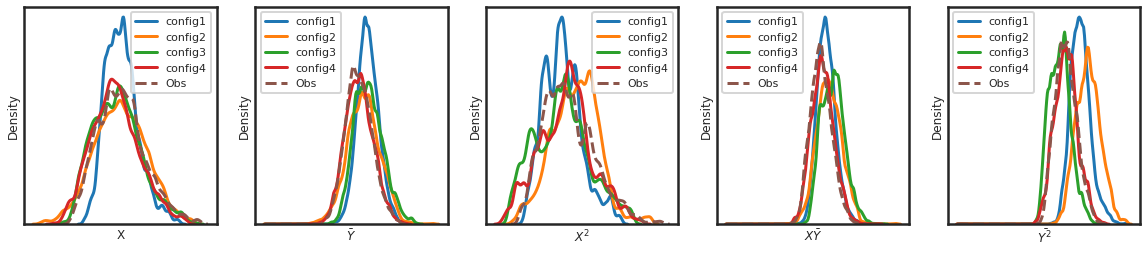}
    \caption{Histograms of metrics on the final 4 configurations identified by HM in experiment with no physical based prior, and the observed metrics.}
    \label{fig:configs_noprior}
\end{figure}

We are thus left with 4 configurations that are equally acceptable reconstructions of the initial L96 simulation (Figure \ref{fig:configs_noprior}). Several methods are available to isolate further a single configuration :
[i] selecting the closest simulation to the observation using for example the median of the five Kullback-Leibler divergences (KL-div) with regard to the observation distribution, [ii] using an ensemble of the simulations and performing a multi-model mean or other weighting schemes. 
Calculating the KL-div for the final 4 configurations yields that configuration \#4 yields the minimal divergence (Table \ref{tableNROY1final}, right hand side column) ; this configuration is indeed the closest to the ground truth configuration. 


As a conclusion from the HM experiment with no physical based prior, we demonstrate that the HM procedure can effectively reconstruct the parameters of the initial L96 simulation, with no physically based prior on the value of these parameters. Notwithstanding, we identify two major challenges that the user is left with. First, the NROY space reduces, to the most, at nearly 2000 configurations of parameters, despite the simplicity of this experiment (which only has 4 tunable parameters, as a reminder). Applying additional statistical techniques and evaluating explicitly some NROY configurations against observations, reduces the number of good candidates for reconstructing the initial simulation to 4. Second, the HM may retain low-implausibility configurations due to the uncertainty in emulators, and not because the metrics are close to observations. Incorporating more physical expertise in HM procedure, as illustrated in the section below, helps resolving these challenges.

\subsection{Incorporating domain expertise in History Matching}
\label{sec:physprior}

In this section, we discuss how important is reducing the prior space, especially if the HM user has a physics-informed initial guess on the value of the parameters. For example, when using the two-scale L96 model, usually the $X$ variables are slower than the $Y$ variables, and their amplitude is higher than those of $Y$. We assume then $c$ and $b$ to be higher than 1. In addition,  we can also introduce knowledge from the domain of application, for example assuming that $F$ and $h$ are positive (which is the case in all studies where L96 model is used as a toy model for climate).  The advantage of HM is that if these new priors do no contain the ground truth solution, the first wave will result in an empty NROY, and the modeler will be enforced to change these priors. 

\begin{table}[H]
\caption{Prior intervals for the parameters for the HM experiment with domain expertise.}
 \centering
\begin{tabular}{c c c}
 \hline
 Params & Prior & True  \\
 \hline
F & {[0,20]} & 10 \\
h & {[0,2]} & 1 \\
c & {[1,20]} & 10 \\
b & {[1,20]} & 10 \\
\hline
\end{tabular}
\label{table2prior}
\end{table}

Performing the HM procedure with physically-informed priors (as listed in Table \ref{table2prior}), required only 5 waves and a total of 200 simulations to reach the same reduction in parameter space as with no physical-based prior (Table \ref{tableNROY2}). Actually, the NROY space becomes less than 1\% of the initial space at wave 4, while it took one additional wave in the previous experiment. While running two waves ($40\times2 = 80$ L96 simulations) is computationally cheap in our study, we highlight that each wave that we would run for a coupled climate model would be very expensive. This illustrates how beneficial it is to chose the prior intervals as carefully as possible, incorporating as much as possible the knowledge from experienced modelers in the domain of application. 

\begin{table}[h]
\label{tableNROY2}
\caption{Same as Table \ref{tableNROY1} for the HM experiment with prior intervals for the parameters limited by domain expertise.} 
\centering
\begin{tabular}{rrr}
  \hline
wave & NROY (\%) & NbSim \\ 
  \hline
1 & 25.57 & 40 \\ 
  2 & 3.18 & 40 \\ 
  3 & 0.59 & 40 \\ 
  4 & 0.04 & 40 \\ 
  5 & 0.02 & 40 \\ 
   \hline & & Total = 200 \\
  \hline
\end{tabular}
\end{table}

\begin{figure}
    \centering
    \includegraphics[scale=0.37]{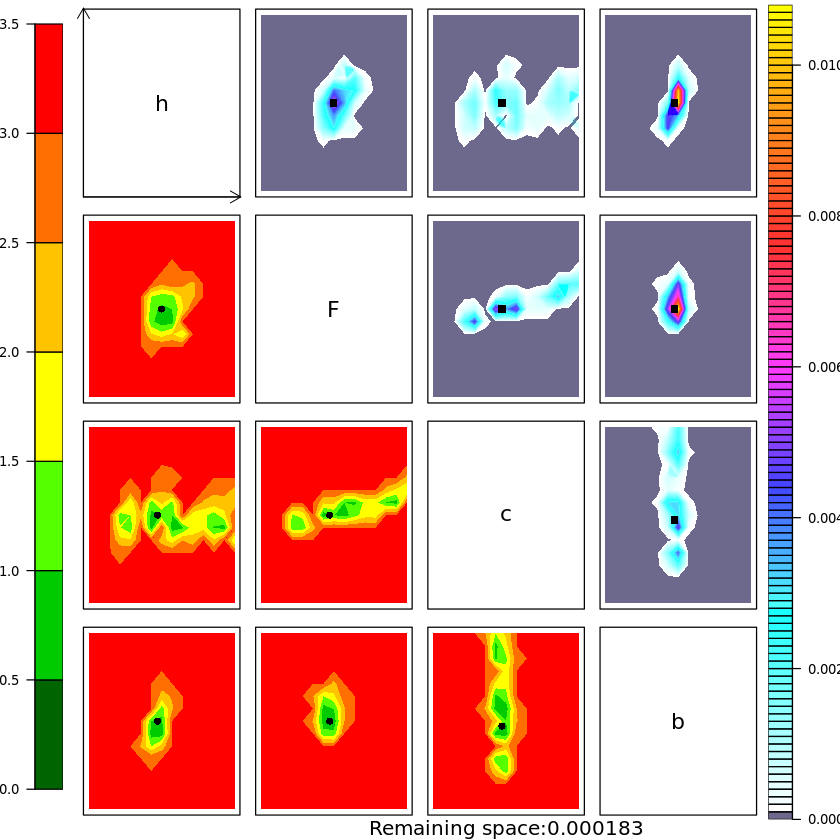}
    \caption{Same as Figure \ref{fig:HM_noprior} for the HM experiment with prior intervals for the parameters limited by domain expertise, at wave 5.}
    \label{fig:HMprior}
\end{figure}

As for the previous experiment, we apply K-means on the final NROY (Figure \ref{fig:HMprior}) and find 3 clusters which centers also belong to the final NROY (Table \ref{configsprior}). 
In this case, all configurations produce metrics which distributions are similar to observations (Figure \ref{fig:configs_Physprior1}).
Interestingly, the remaining 3 configurations have median KL-div that are lower than $0.09$ which is the value for the second best configuration in the previous experiment. This confirms that narrowing the prior helps HM reducing faster the search space and leads to better configurations.

\begin{table}[H]
\centering
\begin{tabular}{rrrrr|c}
  \hline
 & F & h & c & b & KL-div\\ 
  \hline
\textbf{1} & \textbf{9.87} & \textbf{1.05} & \textbf{9.06} & \textbf{10.52} & \textbf{0.03}\\ 
  2 & 10.06 & 0.81 & 11.46 & 9.30 & 0.05\\ 
  3 & 12.13 & 1.11 & 17.10 & 10.48 & 0.07\\ 
   \hline
\end{tabular}
\caption{Same as Table \ref{tableNROY1final} for the HM experiment with domain expertise.}
\label{configsprior}
\end{table}

\subsection{AMIP-style experiment}

Our ambition when applying HM to L96 model is to mimic the calibration of parameters in ocean-atmosphere  models, which couple a fast media, the atmosphere, where variables generally adjust within days to months, and the ocean, the slow media, where some variables require up to hundred years to reach equilibrium. Such calibrating exercise, commonly begins with preliminary calibration of the parameters of each media in uncoupled experiments, which explicitly simulate only one of the two media. In this section, we start with exploring the potential of HM for calibrating the parameters of L96 fast variables alone, i.e. $Y$. As parameter $F$ is not involved directly in the $Y$ variables equation (Eq. \ref{eq:l96Y}), therefore the remaining three parameters are tuned here. Hence at each wave of the HM algorithm, we decide to run 30 simulations only. Those simulations only resolve the equation for $Y$ variables, with $X$ information extracted from the initial simulation, considered as ground truth (see Eq. \ref{eq:l96_AMIP}) :
\begin{gather} \label{eq:l96_AMIP}
    \frac{d Y_{j, k}}{d t}=-\textcolor{blue}{cb} Y_{j+1, k}\left(Y_{j+2, k}-Y_{j-1, k}\right) -\textcolor{blue}{c} Y_{j, k} +\textcolor{blue}{\frac{hc}{b}} \underbrace{ X_{k}}_{\substack{\text{Retrieved } \\ \text{from initial } \\ \text{simulation}}}
\end{gather}
where the parameters to be tuned are highlighted in blue color.

In AMIP-style experiment, the metrics reduce to those concerning $Y$ variables only, i.e. 
\begin{equation} \label{eq:metricsAMIP}
\boldsymbol{f}(p)=\left(\begin{array}{c}
\langle \bar{Y} \rangle_{T} \\
\langle \bar{Y^{2}}\rangle_{T}
\end{array}\right),
\end{equation}

\begin{table}[H]
\label{tableNROYamip}
\caption{Same as Table \ref{tableNROY1} for the AMIP-style HM experiment, where only the fast component of L96 is calibrated. } 
\centering
\begin{tabular}{rrr}
  \hline
 & NROYs & NbSim \\ 
  \hline
1 & 71.86 & 30 \\ 
  2 & 12.19 & 30 \\ 
  3 & 1.24 & 30 \\ 
  4 & 0.12 & 30 \\ 
  5 & 0.01 & 30 \\ 
  \hline & & Total = 150 \\
   \hline
\end{tabular}
\end{table}

Although the NROY after wave 1 is much larger than for previous experiments, only 5 waves are necessary to reduce the NROY to less than $0.01$ \% of the initial parameter space (Table \ref{tableNROYamip}). Still, at wave 5, there remains some uncertainty in the calibration for parameter $c$, as indicated by the band-shape features in $c-$ related panels (Figure \ref{fig:configs_amip}). As a result, the K-means algorithms identifies two solutions, which have very similar KL-div (Table \ref{table:amipfinal}). We conclude that both sets of parameters can be considered as acceptable configurations. Indeed, running simulations with these two sets of parameters yields distributions of metrics that are very close to observations, equally (Figure \ref{fig:configs_amip}). That AMIP-style HM experiment on L96 model returns 2 plausible sets of parameters, reproducing in an equally acceptable manner the single initial simulation, is somehow puzzling. This constitutes one of the lessons we learned from tuning the L96 model, which we discuss in section 4.

\begin{table}[H]
\centering
\begin{tabular}{rrrr|c}
  \hline
 & h & c & b & KL-div\\ 
  \hline
1 & 1.00 & 10.26 & 10.03 & 0.0018\\ 
  2 & 1.05 & 16.98 & 10.41 & 0.0024\\ 
   \hline
\end{tabular}
\caption{Same as Table \ref{tableNROY1final} for the AMIP-style HM experiment.}
\label{table:amipfinal}
\end{table}

\begin{figure}[H]
    \centering
    \includegraphics[scale=0.37]{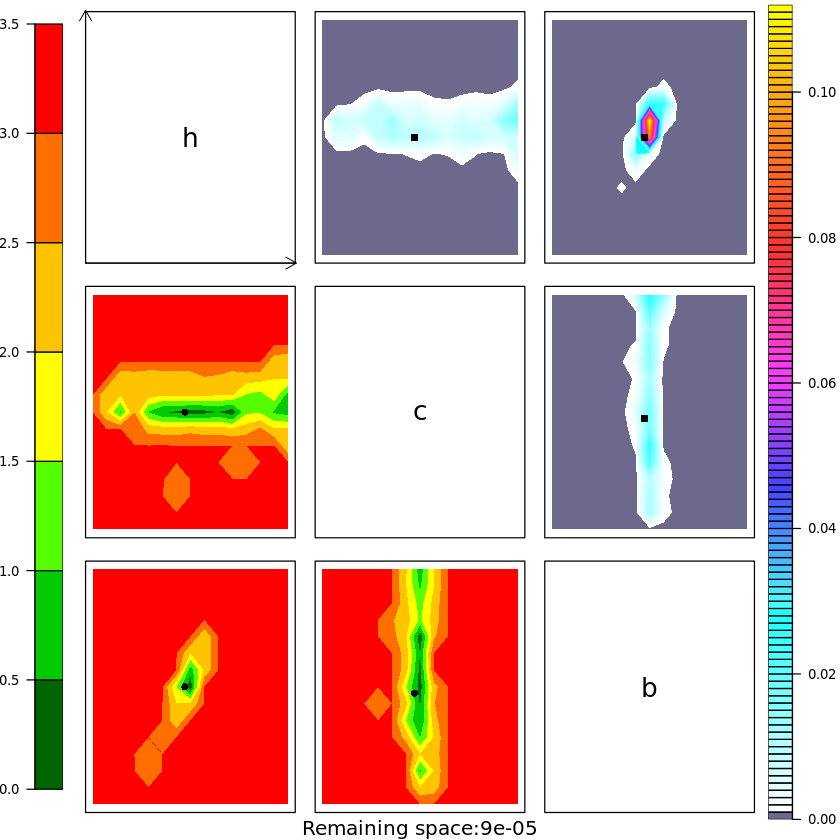}
    \caption{Same as Figure \ref{fig:HM_noprior} for the AMIP-style HM experiment, at wave 5.}
    \label{fig:waveAMIP}
\end{figure}

\begin{figure}
    \centering
    \includegraphics[scale=0.25]{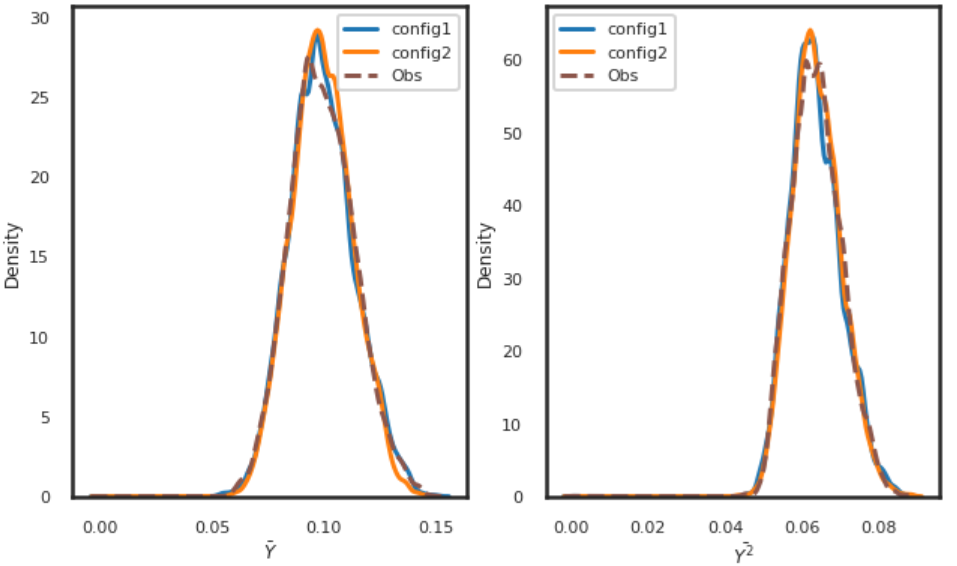}
    \caption{Histograms of metrics on the final 2 configurations identified by AMIP-style HM, and the observed metrics.}
\label{fig:configs_amip}
\end{figure}

\subsection{OMIP-style experiment}

We then explore the symmetric of AMIP-style experiment and  calibrate the parameters affecting L96 slow variables alone, ie $X$. We are left with only two independent parameters to tune, now labeled $F$ and $G$ (Eq. \ref{eq:l96_OMIP}). 

\begin{gather} \label{eq:l96_OMIP}
   \frac{d X_{k}}{dt}=-X_{k-1}\left(X_{k-2}-X_{k+1}\right) -X_{k} + \textcolor{blue}{F} - \textcolor{blue}{G} \underbrace{\sum_{j=1}^{J} Y_{j, k}}_{\substack{\text{Retrieved from} \\ \text{initial simulation}}} 
\end{gather}

\begin{figure}
    \centering
    \includegraphics[scale=0.37]{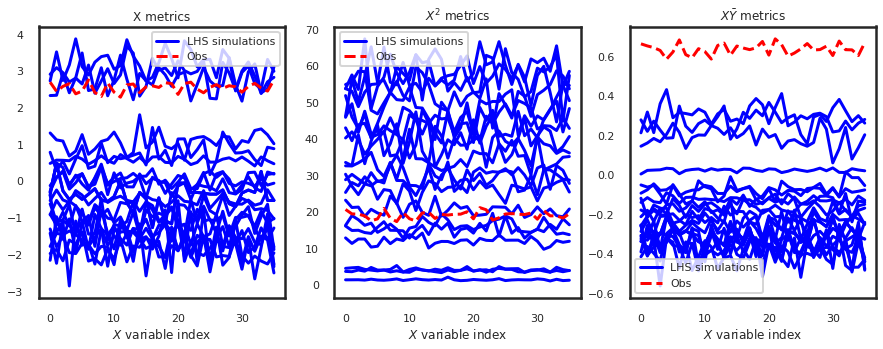}
    \caption{Non-dimensional metrics for the OMIP-style experiment : for the ground truth (red lines) and for a selection of parameters (blue lines).}
    \label{fig:metricsOMIP}
\end{figure}

The prior intervals for both parameters is the same as for previous experiments, $[0,20]$, and we initially include all metrics that concern the $X$ variables, ie 
\begin{equation} \label{eq:metricsOMIP}
\boldsymbol{f}(X, Y)=\left(\begin{array}{c}
\langle X\rangle_{T} \\
\langle X^{2}\rangle_{T} \\
\langle X \bar{Y}\rangle_{T}
\end{array}\right)
\end{equation}
With such definition of metrics, the NROY at wave 1 is empty, which seems to be due to mismatches between the initial simulation and the OMIP-style outputs for $\langle X \bar{Y}\rangle_{T}$ metrics primarily (Figure \ref{fig:metricsOMIP}). This is presumably due to the lack of coupling between $X$ and $Y$ variables when running L96 in OMIP-style experiment (cf Eq. \ref{eq:l96_OMIP}). Actually, $\langle X \bar{Y}\rangle_{T}$  metrics, which associates the slow and fast variables of L96 model, stands for any properties emerging from the coupling of an ocean with an atmosphere, for example sea-ice related quantities. 

When removing the $\langle X \bar{Y}\rangle_{T}$ metric from the evaluation against ground truth in OMIP-style experiment, then a single wave of HM reduces the parameter space to $1.8$\%. The final NROY space includes a single cluster centered on values $[F,G]=[10.23, 1.06]$, which are very close to the ground truth. As a conclusion, HM is efficient in OMIP-style experiments only if metrics resume to those affecting the slow variable exclusively.

\begin{figure}[H]
    \centering
    \includegraphics[scale=0.37]{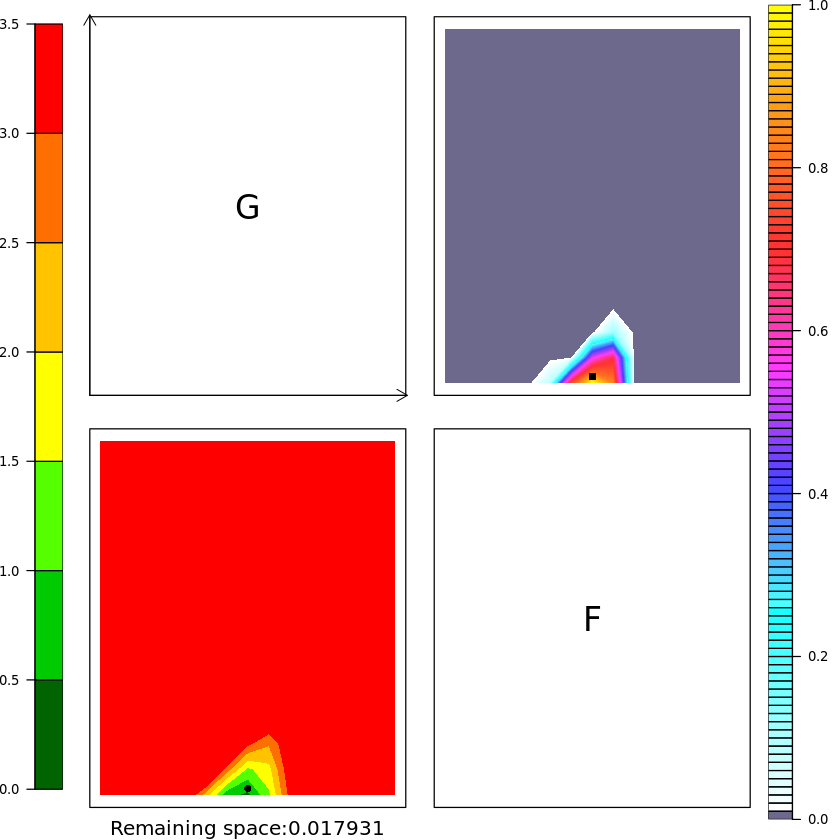}
    \caption{OMIP tuning experiment: NROY plot after one wave}
    \label{fig:waveOMIP}
\end{figure}


\section{Discussion}
\label{sec:discussion}

The Lorenz96 system, since it was first articulated by Lorenz, has been used to understand general circulation models from a dynamical systems perspective. In its typical use (such as by Lorenz himself in \cite{lorenz1996predictability}), the $X$ layer stands in for the resolved-scale flow, and the $Y$ for subgrid-scale unresolved flows, to study the response of the full system to the aggregate behavior of the unresolved scales. 
In this manuscript we take the same L96 system to study a different aspect of the coupled climate system, with a fast ($Y$, ``atmosphere'') and slow ($X$, ``ocean'') component, with the coupling representing the surface exchanges. 

Our objective is to revisit the problem of model calibration in the presence of distinct timescales in separate components of the coupled system. As noted in studies of model calibration or ``tuning'' \citep[e.g.][]{ref:hourdinetal2017}, this is generally done by first separately calibrating individual components of the climate system to conform to theoretical and observational constraints, and then in a separate step, applying global constraints on emergent properties of the coupled climate system, such as top-of-atmosphere radiative balance, or ENSO behavior \citep{ref:heldetal2019, ref:mignotetal2021}. The initial stages are done with fixed surface exchanges: for example in AMIP experiment,  \cite{ref:zhaoetal2018a} calibrates the atmosphere under prescribed sea surface temperatures. As the intent is not to ``retune'' the atmosphere after coupling, there is a careful assessment of how the ocean might react, by looking at diagnostics such the ``implied'' ocean heat transport \citep{ref:zhaoetal2018b}. The ocean is likewise first calibrated using fixed atmospheric forcings, following the OMIP protocol described in \cite{ref:griffiesetal2016}. We find however, that we must then refine the ocean tuning in a coupled setting \citep{ref:adcroftetal2019}, to adjust the tunings to match the differences between the standard atmosphere forcings of AMIP and the actual atmospheric model to be used in the GCM. This carries the risk, of course, of compensating for atmospheric model biases by adjusting ocean parameters. When tuning the IPSL climate model for CMIP6, it was actually intentional to compensate for biases in the atmosphere, that initially limited dense water formation, by tuning the sea-ice leads fraction and finally enhance ventilation of the deep ocean \citep{ref:mignotetal2021}.

As GCMs move toward more objective methods of model calibration, we must examine these nuances introduced by coupling. Here we have emulated the HM-based model calibration process in the L96 system. An initial ``truth'' run of L96 serves as the history matching target for a given set of values of the parameters $\{F,h,c,b\}$: the aim is to discover optimal tuning values of these parameters from a given prior distribution of possible values. This is thus a ``perfect model'' setting, and only parametric error can be studied, not structural error, since the same equations are used in the test model, with only the parameters to be discovered. In an ocean-atmosphere coupled model, as well as in a more complex climate model, there are structural biases inherent to the grid resolution, the choice of algorithms for discretization, and/or poorly known physics, in particular at the interface between the different components. The HM procedure enables us to factor those in through uncertainty in the metrics, but an accurate quantification of those uncertainties is not straightforward as it often combines with parametric uncertainties \citep{balaji2022gcms}. Also when tuning a climate model, metrics refer to observations, that may encompass intrinsic uncertainty which should be taken into account, as well, this limitation cannot be envisioned in a perfect model setting. 

When tuning L96 model, we use the set of metrics shown in Eq.~\ref{eq:metrics}: they are first- and second-order metrics covering the slow and fast components separately, as well as a cross-term metric for the emergent properties of the coupled system. In calibrating a real GCM, we of course consider more complex and higher-order metrics, but broadly they will fall into one of these categories: fast, slow and emergent-coupled. It must be noted however that the L96 system is extremely simple compared to real GCMs, leading to some limitations in the applicability of our conclusions. Notwithstanding, our conclusion that ocean-only experiments cannot be used to calibrate ocean parameters if the emergent-coupled metrics are included in the HM procedure, seems to be robust. For example, tuning sea-ice quantities in IPSL-CM6A-LR had to be done in coupled mode \citep{ref:mignotetal2021}. This advocates for running OMIP experiments to calibrate ocean quantities related to the ocean interior dynamics, and calibrating upper-ocean variables, including sea-ice, in coupled mode. 

A second aspect of objective calibration that we consider, is the independence of metrics. As there are many measures of validation, objective methods can rapidly become expensive if they are all to be taken into account. However there is considerable covariance between metrics. We explore here the use of principal component analysis to find a minimal optimal set of metrics that capture the variance across metric space, similar to the approach in \cite{ref:bakeretal2015}. Even in the case of L96 model with a relatively small number of metrics as compared to a climate model, it appears efficient to reduce dimensionality of the metrics, hence we strongly recommend to proceed similarly with climate models, in line with \cite{salter2019uncertainty}.

Finally, a paramount aspect of HM for parameter calibration is the fact that by ruling out parameters rather than selecting an optimal set, the uniqueness of plausible configurations is revoked. This appears clearly when applying HM in AMIP-style experiments of L96 model : HM procedure identifies 2 sets of parameters that yield equally satisfying metrics, which only 1 is close to ground truth (Table \ref{table:amipfinal}). The non-uniqueness of parameters, for a given model code and specific metrics, is an emergent property of GCMs \citep{ref:hourdinetal2022} that must be discussed in the context of CMIP exercises.

\section{Conclusion and future work}
\label{sec:conclusions}

We find that the L96 system can indeed shed light on certain aspects of the calibration of coupled models with two intrinsic timescales, providing lessons for the use of HM or other objective calibration methods in GCM development. The key findings of this paper can be summarized as follows. 

First, we find that the HM method does indeed allow calibration of models close to optimal values, with minimal errors relative to the true values of $\{F,h,c,b\}$. Simulations with the tuned values of these parameters produce simulations whose ``climate'' -- time-averaged metrics -- is statistically close to the initial simulation considered as ground truth. 3 waves of tuning efficiently reduces the NROY space to a fraction $\order(10^{-3}-10^{-4})$ of the prior volume of parameter space to be explored. Making the system more chaotic -- which can be done by increasing the value of $F$ -- does introduce more noise into the calibration, which can stray further from the true values \citep{ref:christensenberner2019} . 

In the meanwhile, we show that we can reduce the space of metrics considerably by finding the eigenfunctions in metric space that explain most of the variance. Non-linear PCA and ML-based methods of dimensionality reduction do not show much benefit relative to standard PCA, at least in the L96 system. We cannot infer from this that PCA might not be of use for reducing metric space in a real GCM; it has been used for related purposes in other contexts, such as estimating the similarity of climate across simulations \citep{ref:bakeretal2015}.

Second, we illustrate how much the number of waves hence explicit simulations of the L96 model can be drastically reduced if physical expertise is introduced prior to the initiation of HM algorithm. Additionally, the remaining NROY includes a smaller number of clusters, hence plausible sets of parameters to mimic ground truth. This advocates to merging as much physical expertise as possible when setting the  prior intervals for each parameters. It's important to underline here that the goal of HM here is not to find the true value of parameters used for the initial experiment. Rather, it is to find acceptable values that produce a similar climate. Hence it has to be provided that each cluster from the final NROY feeds into simulations, which climate is then compared to the ground truth. 

Third, we demonstrate that the AMIP--OMIP--coupled sequence of tuning is indeed capable of introducing compensating errors into the coupled system, due to metrics emerging from the coupling between the slow and the fast components, such as sea-ice for example. Additionally, we can sometimes find a non-compact NROY with two ``clusters'' with different values of two parameters (Figure \ref{fig:waveAMIP}), that yield equally satisfying metrics (Figure \ref{fig:configs_amip}). Here,  physical insight might play a key role in deciding which NROY cluster to select. One should then be careful when using clustering techniques, k-means has its shortcomings such as the lack of flexibility in cluster shape. For complex (e.g. non-elliptic) clusters, DBSCAN or other density-based clustering techniques can be explored, but then the idea of a representative centroid of the cluster is not relevant anymore. 

The particularity of a potential use of HM for coupled climate models is the difficulty, or even the impossibility, of running hundreds of high cost simulations. Active learning techniques can help overcome the greedy sampling done is the iterative refocussing step. It resorts to choosing the next design point by solving an optimization problem \citep{craig1997pressure}, recent work by \citet{garbuno2020history} delves into this question and presents three active learning criteria. We believe that advancements in this direction might help lower the cost of a global coupled model HM. 

As a final word, as we learned many lessons from applying HM to L96 model in the context of climate model tuning, it appears clearly that HM should not be seen as a fully automatic stand-alone solution for climate model tuning. Rather, it should be seen as a semi-automatic tool that, combined with physical expertise on climate model sensitivities, can provide efficient and robust quantification of the parametric uncertainties. Also, as compared to by-hand tuning, it offers a unique opportunity to reduce the number of climate model simulations yet exploring a larger diversity of parameter sets, which helps transitioning climate modeling to a more sustainable science. Hence we encourage climate modelers to start incorporating it in the tuning road-map of their models, as a first step towards quantifying the overall uncertainty of climate projections.

\appendix

\section{Metrics for the resulting configurations}
\label{allconfigs}

\subsection{Experiment 1: tuning L96 with no prior}

\begin{figure}[H]
    \centering
    \includegraphics[scale=0.32]{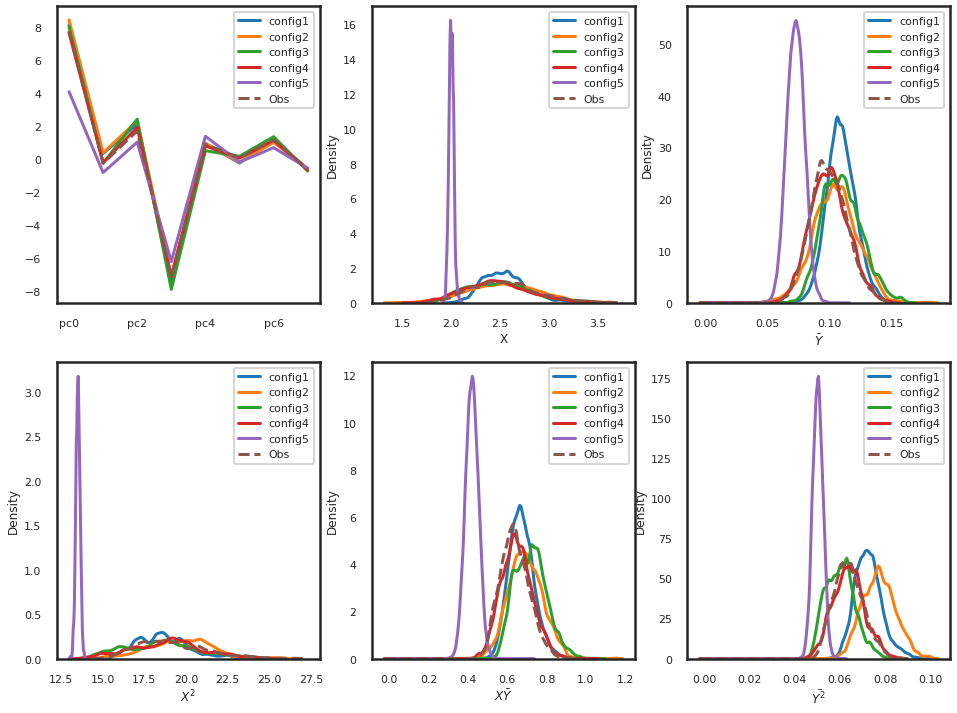}
    \caption{Histograms of the metrics on the configurations (straight lines) and the observed metrics (dashed line)}
    \label{fig:configs_noprior1}
\end{figure}

\subsection{Experiment 2: tuning L96 with physical prior}

\begin{figure}[H]
    \centering
    \includegraphics[scale=0.32]{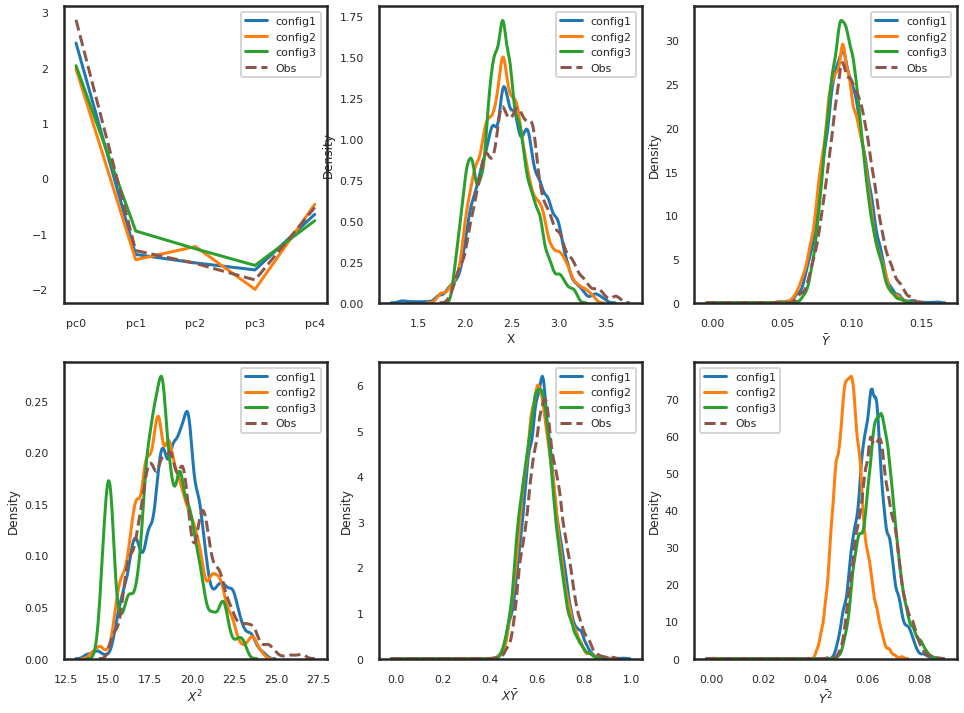}
    \caption{Histograms of the metrics on the configurations (straight lines) and the observed metrics (dashed line)}
    \label{fig:configs_Physprior1}
\end{figure}




\section{Silhouette scores}
\label{silhou}

Presented in this appendix are silhouette scores used for deciding the number of clusters in the K-means clustering

\begin{figure}[H]
    \centering
    \includegraphics[scale=0.3]{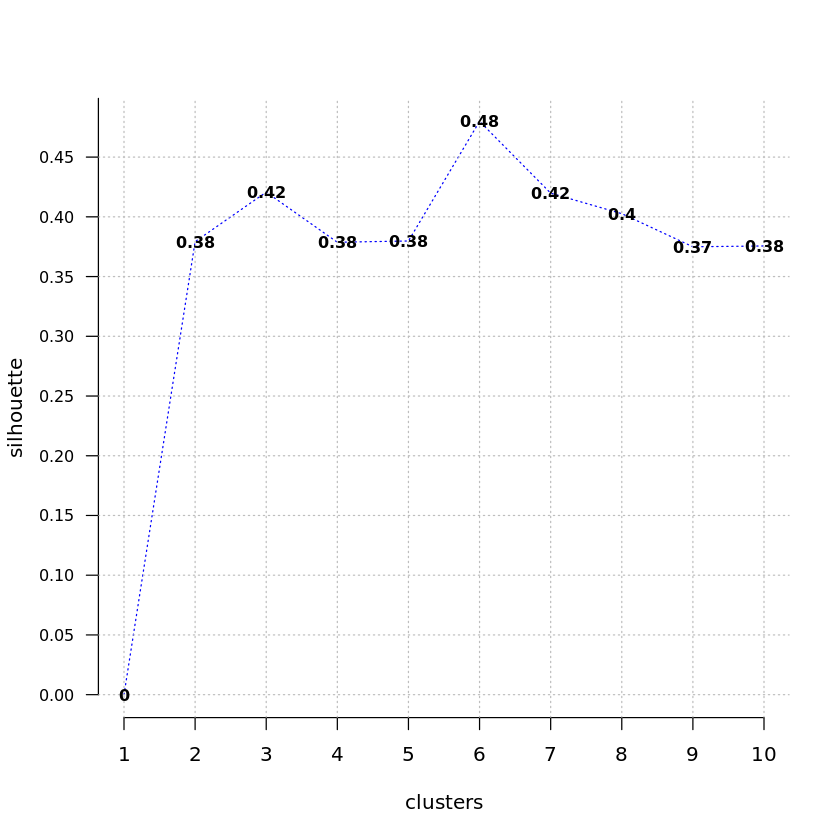}
    \caption{L96 tuning experiment without physical prior: silhouette score, here optimal number of classes is 6}
    \label{fig:silhouette_noprior}
\end{figure}

\begin{figure}[H]
    \centering
    \includegraphics[scale=0.3]{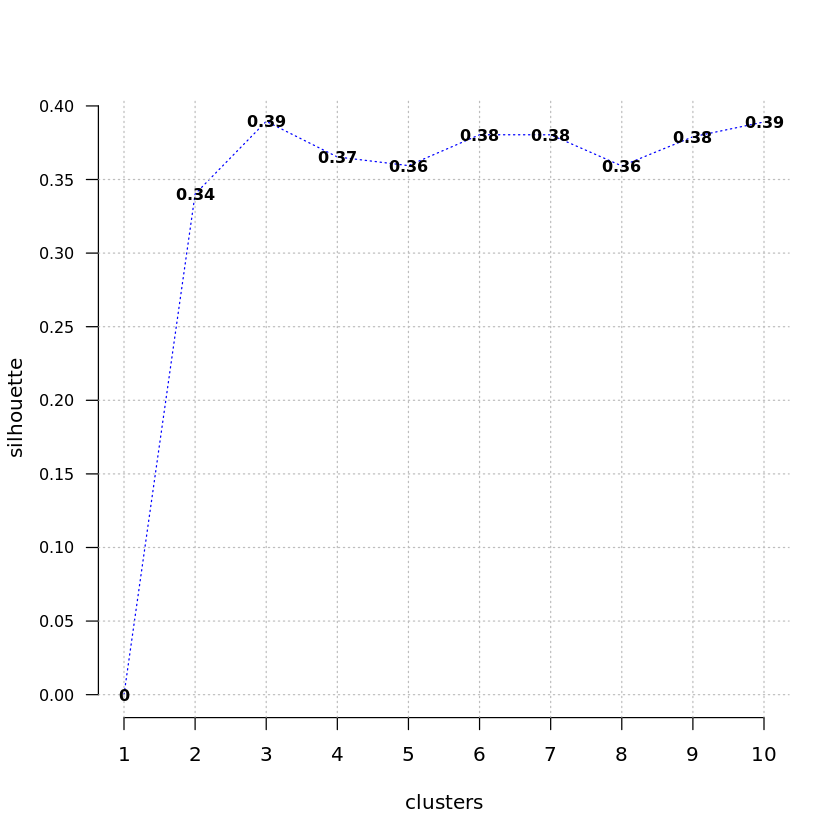}
    \caption{L96 tuning experiment with physical prior: silhouette score, here optimal number of classes is 3}
    \label{fig:silhouette_noprior}
\end{figure}

\begin{figure}[H]
    \centering
    \includegraphics[scale=0.3]{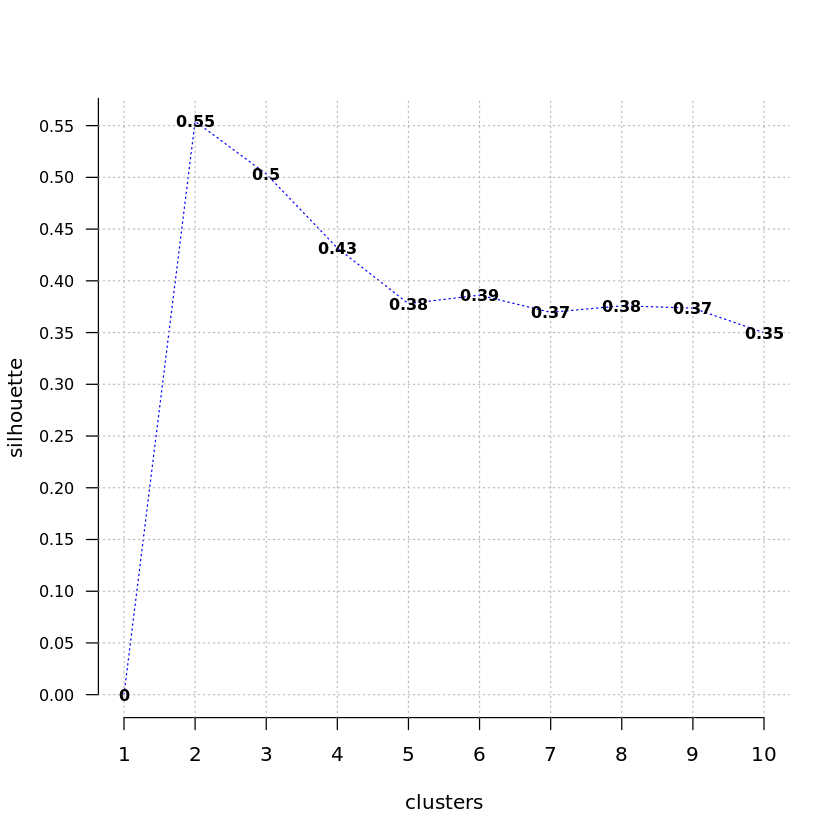}
    \caption{AMIP experiment: silhouette score, here optimal number of classes is 2}
    \label{fig:silhouette_noprior}
\end{figure}

\acknowledgments
The authors would like to thank Frederic Hourdin, Fleur Couvreux, Daniel Williamson, Najda Villefranque and Martin Vancoppenolle for the helpful discussions. This research was supported by the "Agence Nationale de la Recherche" through the HRMES ANR-17-MPGA-0010 project. This work was granted access to the HPC/AI resources of IDRIS under the
allocation A0120107451 made by GENCI.


%
%

\bibliography{agusample,brefs}

%
%
%
%
%

\end{document}